\documentclass[twoside,fleqn]{article}
\usepackage{espcrc2,graphicx,epsfig,cite,amssymb}

\newcommand{\be}{\begin{equation}}
\newcommand{\ee}{\end{equation}}
\newcommand{\ba}{\begin{eqnarray}}
\newcommand{\ea}{\end{eqnarray}}
\newcommand{\chpt}{$\chi$PT}
\newcommand{\cO}{{\cal O}}
\newcommand{\cJ}{{\cal J}}

\newcommand{\AmS}{{\protect\the\textfont2
  A\kern-.1667em\lower.5ex\hbox{M}\kern-.125emS}}

\title{Chiral low-energy constants $\mathbf{L_{10}}$ and $\mathbf{C_{87}}$ from hadronic\ {\Large $\mathbf{\tau}$}\ decays 
}

\author{Mart\'{\i}n Gonz\'alez-Alonso$^a$\thanks{Speaker. E-mail: Martin.Gonzalez@ific.uv.es.},
Antonio Pich$^a$ and Joaquim Prades$^b$
\\ [0.4cm]
${}^a$
Departament de F\'{\i}sica Te\`orica and IFIC, Universitat de Val\`encia-CSIC,\\
Apt. Correus 22085, E-46071 Val\`encia, Spain\\
${}^b$ CAFPE and Departamento de F\'{\i}sica Te\'orica y del Cosmos,\\
Universidad de Granada, Campus de Fuente Nueva, E-18002 Granada,
Spain}

\begin{document}
\begin{abstract}
Using recent precise hadronic $\tau$-decay data on the $V\!-\!A$ spectral function and general properties of QCD such as analyticity, the operator product expansion (OPE) and chiral perturbation theory (\chpt), we  get accurate values for the QCD chiral order parameters $L_{10}^r$ and $C_{87}^r$. At order $p^4$ we obtain $L_{10}^r(M_\rho) = -(5.22\pm 0.06)\cdot 10^{-3}$, whereas at order $p^6$ we get $L_{10}^r(M_\rho) = -(4.06\pm 0.39)\cdot 10^{-3}$ and $C_{87}^r(M_\rho) = (4.89\pm 0.19)\cdot 10^{-3}\;\mathrm{GeV}^{-2}$.
\end{abstract}
%
\maketitle
\section{Introduction}
Hadronic $\tau$-decay data are a very important source of information, both on perturbative and non-perturbative QCD. Of special interest in order to study non-perturbative QCD quantities is the difference of the vector and axial-vector spectral functions, because in the chiral limit the corresponding $V\!-\!A$ correlator is exactly zero in perturbation theory.

The $\tau$ data can be used to determine the parameters of \chpt\ \cite{WEI79}, the effective field theory of QCD at very low energies (a Taylor expansion in external momenta and quark masses). At lowest order, ${\cal O}(p^2)$, the SU(3) \chpt\ Lagrangian has only two parameters, the pion decay constant $f_\pi$ and the light quark condensate. At ${\cal O}(p^4)$ twelve more low-energy constants (LECs) appear ($L_{i=1,\cdots,10}$ and $H_{1,2}$), whereas at ${\cal O}(p^6)$ we have 90 (23) additional parameters $C_i$ in the even (odd) intrinsic parity sector\cite{p6}. These LECs are related to order parameters of the spontaneous chiral symmetry breaking of QCD, and have to be determined phenomenologically or using non-perturbative techniques. For the $L_i$ couplings this has been done to an acceptable accuracy, but the $C_i$ LECs are less well known.

There has been a lot of recent activity to determine these chiral LECs from theory, using as much as possible QCD information \cite{MP08,MOU97,KN01,CEE04,CEE05,KM06,PRS08}. This strong effort is motivated by the precision required in present phenomenological applications, which makes necessary to include corrections of ${\cal O}(p^6)$. The huge number of unknown couplings is the major source of theoretical uncertainty.

We present here an accurate determination of the \chpt\ couplings $L_{10}$ and $C_{87}$ \cite{GPP08}, using the most recent hadronic $\tau$-decay data \cite{ALEPH05}. Estimates of $L_{10}$ from $\tau$-data have been done previously \cite{DHG98,NAR01,DS}, altough our analysis is the first that includes the known two-loop \chpt\ contributions and then also the first that provides $C_{87}$.
\section{Theoretical Framework}
The basic objects of the theoretical analysis are the two-point correlation functions of the non-strange vector and axial-vector quark currents
\ba
\label{eq:two}
\Pi^{\mu\nu}_{ij,\cJ}(q)
\equiv i \!\int\!\! \mathrm{d}^4 x \,  \mathrm{e}^{i q x}
\langle 0 | T \!\left( \cJ_{ij}^\mu(x) \cJ_{ij}^\nu(0)^\dagger \right)\!| 0 \rangle \nonumber \\
\!\!\!\!=\! (-\!g^{\mu\nu}\! q^2 \!+\! q^\mu\! q^\nu ) \Pi^{(1)}_{ij,\cJ}(q^2) + q^\mu\! q^\nu \Pi^{(0)}_{ij,\cJ}(q^2) ,\!\!
\ea
where $\cJ_{ij}^\mu$ denotes $V_{ud}^\mu\!=\!\overline{u} \gamma^\mu d$ and $A_{ud}^\mu\!=\!\overline{u} \gamma^\mu \gamma_5 d$.
In particular we are interested in the difference $\Pi(s) \equiv \Pi_{ud,V}^{(0+1)}-\Pi_{ud,A}^{(0+1)}$, and we will work in the isospin limit ($m_u=m_d$) where $\Pi^{(0)}_{ud,V}(q^2)=0$.

From the analytic structure of the correlator $\Pi(s)$ in the complex $s$-plane and its OPE one can get the
following two sum rules (see ref.~\cite{GPP08} for a careful derivation)
\ba
\label{eq:SR1}
-8 \, L_{10}^{\rm eff} \!\!\!\!&\equiv&\!\!\!\! \int^{s_0}_{s_{\rm th}} \frac{\mathrm{d}s}{s} \frac{1}{\pi} \, {\rm Im} \, \Pi(s)
= \frac{2 f_\pi^2}{m_\pi^{2}} +\! \Pi(0) \\
\label{eq:SR2}
16 \, C_{87}^{\rm eff} \!\!\!\!&\equiv&\!\!\!\! \int^{s_0}_{s_{\rm th}} \frac{\mathrm{d}s}{s^2} \frac{1}{\pi} \, {\rm Im} \, \Pi(s)
= \frac{2 f_\pi^2}{m_\pi^{4}} +\! \frac{\rm{d}\Pi}{\rm{ds}}(0)~,
\ea
that represent the starting point of our work. The interest of these two relations stems from the fact that the effective parameters $L_{10}^{\rm eff}$ and $C_{87}^{\rm eff}$ can be extracted from the data and the r.h.s can be rigorously calculated within \chpt~in terms of the LECs that we want to determine. From the results of ref.~\cite{ABT00} we get
\ba
\label{L10-p6}
\lefteqn{2 f_\pi^2 / m_\pi^{2} +\Pi(0)=-8L_{10}^r(\mu) + G^4_{1L}(\mu)}  \nonumber\\
&&  +\,\, G^6_{0L}(\mu) + G^6_{1L}(\mu) + G^6_{2L}(\mu) +\cO(p^8)\nonumber\\
\label{C87-p6}
\lefteqn{2 f_\pi^2 / m_\pi^4 + \Pi{}'(0) = H^4_{1L}} \nonumber\\
&& +\,\, 16\,C_{87}^r(\mu)+H^6_{1L}\!(\mu)+H^6_{2L}\!(\mu) + \cO(p^8)~\!,
\ea
where the functions $G^m_{nL}(\mu), H^{m}_{nL}(\mu)$ are corrections of order $p^m$ generated at the $n$-loops level. We omit their
explicit analytic form \cite{GPP08} for simplicity, but it is important to say that $G^6_{0L,1L}(\mu)$ contain some LECs that will represent
the main source of uncertainty for $L_{10}^r$.
\section{Determination of Effective Couplings}
We will use the recent ALEPH data on hadronic $\tau$ decays \cite{ALEPH05}, that provide the most precise measurement of the $V\!\!-\!A$ spectral function.

The relations (\ref{eq:SR1}) and (\ref{eq:SR2}) are exactly satisfied only at $\rm{s_0\!\to\!\infty}$, but we are forced to take finite values of $s_0$ neglecting in this way the rest of the integral\footnote{Equivalently, we are assuming that the OPE is a good approximation for $\Pi(s)$ at any $|s|\!\!=\!\!s_0$, what is not expected to happen near the real axis and that produces the DV.}. From the $s_0$-sensitivity of the effective parameters one can assess the size of this theoretical error (quark-hadron duality violation -DV-).

\begin{figure}[tbh]
\vfill

\centerline{
\begin{minipage}[t]{.3\linewidth}\centering
\centerline{\includegraphics[width=7cm]{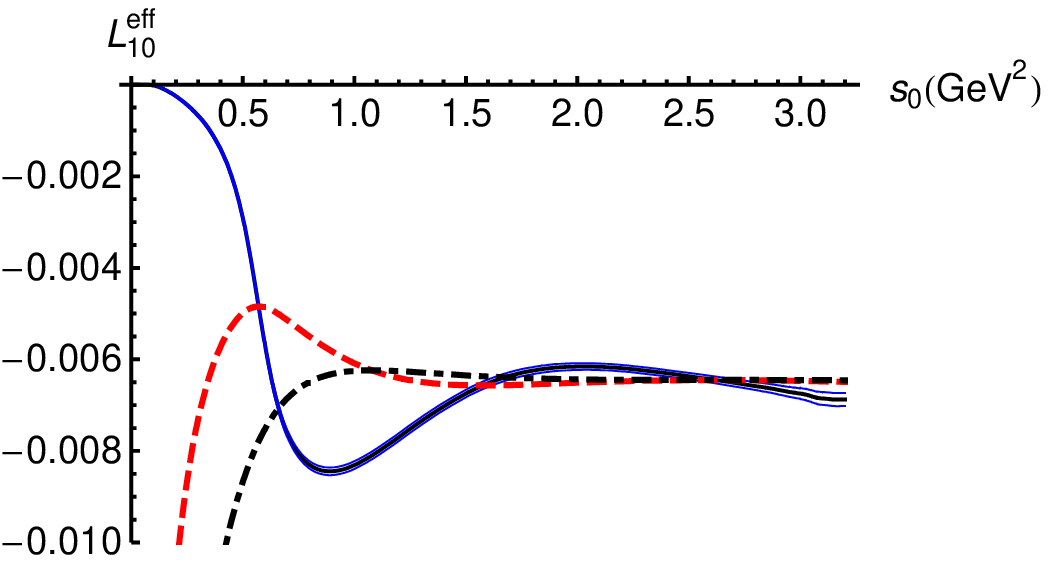}}
\end{minipage}
}

\vspace{0.67cm}
\centerline{
\begin{minipage}[t]{.3\linewidth}\centering
\centerline{\includegraphics[width=7cm]{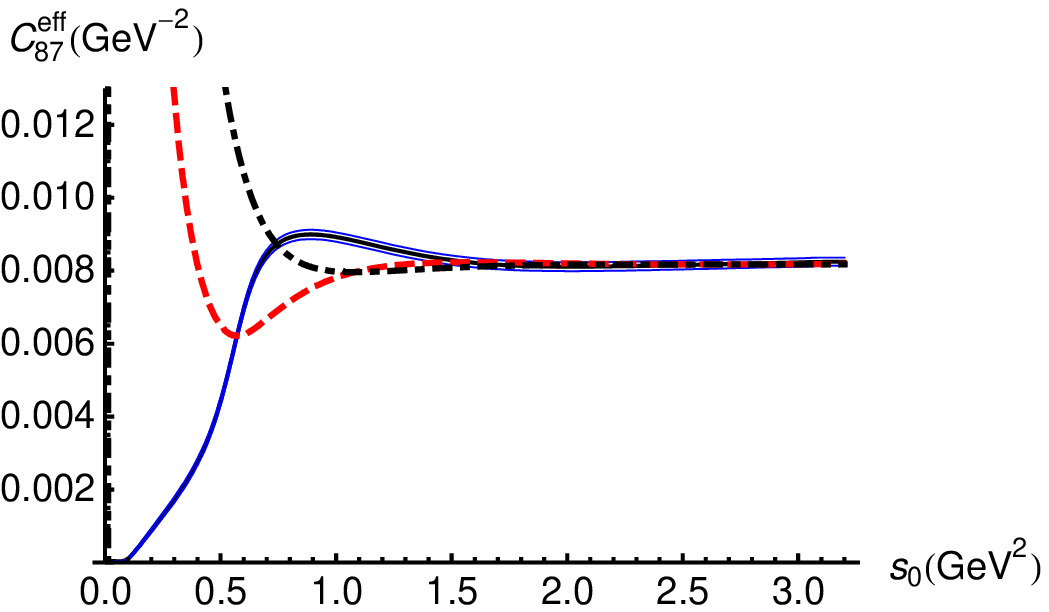}}
\end{minipage}
}

\vfill
\caption{$L_{10}^{\rm eff}(s_0)$ and $C_{87}^{\rm eff}(s_0)$ from different sum rules. For clarity, we do not include the error bands associated with the modified weights.}
\label{fig:L10C87}
\end{figure}

In Fig. \ref{fig:L10C87}, we plot the value of $L_{10}^{\rm eff}$ obtained for different values of $s_0$, with the one-sigma experimental error band, and we can see a quite stable result at $s_0\!\gtrsim\! 2~\mathrm{GeV}^2$ (solid lines). The weight function $1/s$ decreases the impact of the high-energy region, minimising the DV; the resulting integral appears then to be much better behaved than the sum rules with $s^n$ ($n\ge0$) weights.
%

There are some possible strategies to estimate the value of $L_{10}^{\rm eff}$ and his error. One is to give the predictions fixing $s_0$ at the so-called ``duality points'', two points where the first and second Weinberg sum rules (WSR) \cite{WSR} happen to be satisfied. In this way we get $L_{10}^{\rm eff} = -(6.50\pm 0.13) \cdot 10^{-3}$, where the uncertainty covers the values obtained at the two ``duality points''.

If we assume that the integral (\ref{eq:SR1}) oscillates around his asymptotic value with decreasing oscillations and we perform an average between the maxima and minima of the oscillations we get $L_{10}^{\rm eff} = -(6.5\pm 0.2) \cdot 10^{-3}$.

Another way of estimating the DV uses appropriate oscillating functions defined in \cite{GON07} which mimic the real quark-hadron oscillations above the data. These functions are defined such that they match the data at $\sim\!3~\mathrm{GeV}^2$, go to zero with decreasing oscillations and satisfy the two WSRs. We find in this way $L_{10}^{\rm eff} = -(6.50\pm 0.12) \cdot 10^{-3}$, where the error spans the range generated by the different functions used.

Finally we can take advantage of the WSRs to construct modified sum rules with weight factors $w(s)$ proportional to $(1-s/s_0)$, in order to suppress numerically the role of the suspect region around $s\!\sim\!s_0$ \cite{LDP:92}. Fig.~\ref{fig:L10C87} shows the results obtained with $w_1(s)\!\equiv\!\left(1\!-\!s/s_0\right)/s$ (dashed line) and $w_2(s)\!\equiv\!\left(1\!-\!s/s_0\right)^2/s$ (dot-dashed line). These weights give rise to very stable results over a quite wide range of $s_0$ values. One gets $L_{10}^{\rm eff} = -(6.51\pm 0.06) \cdot 10^{-3}$ using $w_1(s)$ and $L_{10}^{\rm eff} = -(6.45\pm 0.06) \cdot 10^{-3}$ using $w_2(s)$.

Taking into account all the previous discussion, we quote as our final conservative result:
\be
\label{L10eff}
L_{10}^{\rm eff} = -(6.48\pm 0.06) \cdot 10^{-3} \, .
\ee
We have made a completely analogous analysis to determine $C_{87}^{\rm eff}$. The results are shown in Fig.~\ref{fig:L10C87}. The solid lines, obtained from Eq.~(\ref{eq:SR2}), are much more stable than the corresponding results for $L_{10}^{\rm eff}$, due to the $1/s^2$ factor in the integrand. The dashed and dot-dashed lines have been obtained with the modified weights $\rm{w_3(s)\equiv\! \frac{1}{s^2}\!\left(1\!-\!\frac{s^2}{s_0^2}\right)}$ and $\rm{w_4(s)\equiv\! \frac{1}{s^2} \left(1\!-\!\frac{s}{s_0}\right)^2 \left(1\!+\!2\frac{s}{s_0}\right)}$. The agreement among the different estimates is quite remarkable, and our final conservative result is
\be
\label{C87eff}
C_{87}^{\rm eff} = (8.18\pm 0.14) \cdot 10^{-3} \, {\rm GeV}^{-2} \, .
\ee
\section{Determination of $L_{10}^r$ and $C_{87}^r$ }
\label{determination}
The \chpt\ coupling $L_{10}^{r}(\mu)$ can be obtained from $L_{10}^{\rm eff}$, using the relation (\ref{L10-p6}). At $\cO(p^4)$ the determination is straightforward and one gets
\be
\label{valL10p4}
L_{10}^r(\mu\!=\!M_\rho) = -(5.22 \pm 0.06) \cdot 10^{-3}  \, .
\ee
At order $p^6$, the numerical relation is more involved because it gets small corrections from other LECs. It is useful to classify the $\cO(p^6)$ contributions through their ordering within the $1/N_C$ expansion. The tree-level term $G_{0L}^6(\mu)$ contains the only $\cO(p^6)$ correction in the large--$N_C$ limit, $\rm{4 m_\pi^2 (C_{61}^r \!-\! C_{12}^r \!-\! C_{80}^r)}$, that is numerically small because of the $m_\pi^2$ suppression and can be estimated with a moderate accuracy \cite{KM06,JOP04,UP08,CEE05,ABT00}.

At NLO $G_{0L}^6(\mu)$ contributes with a term of the form $m_K^2(C_{62}^r \!-\! C_{13}^r \!-\! C_{81}^r)$. In the absence of information about these LECs we will adopt the conservative range $|C_{62}^r \!- C_{13}^r \!- C_{81}^r\!| \le |C_{61}^r \!- C_{12}^r \!- C_{80}^r|/3$, which generates the uncertainty that will dominate our final error on $L_{10}^r$. Also at this order in $1/N_C$ there is the one-loop correction $G_{1L}^6(\mu)$ that is proportional to $L_{9}^r$, which is better known \cite{BT02}.
Calculating the $1/N_C^2$ suppressed two-loop function $G_{2L}^6(\mu)$ and taking all these contributions into account we finally get the wanted $\cO(p^6)$ result:
\ba
\label{valL10p6}
L_{10}^r(M_\rho) \!\!\!&=&\!\!\! -(4.06 \pm 0.04_{L_{10}^{\mathrm{eff}}}\pm 0.39_{\mathrm{LECs}}) \cdot 10^{-3} \nonumber\\
		 \!\!\!&=&\!\!\! -(4.06 \pm 0.39) \cdot 10^{-3} \, .
\ea
where the error has been split into its two main components.
Repeating the same process with $C_{87}^r$ (where the only LEC involved is $L_9^r$) we get
\be
\label{valC87}
C_{87}^r(M_\rho) = (4.89 \pm 0.19) \cdot 10^{-3} \:\mathrm{GeV}^{-2}\, .
\ee
\section{Summary}
Using general properties of QCD and the measured $V\!-\!A$ spectral function \cite{ALEPH05} we have determined the chiral LECs $L_{10}^r(M_\rho)$ and $C_{87}^r(M_\rho)$ rather accurately, with a careful analysis of the theoretical uncertainties.

There are other determinations of $L_{10}$ from $\tau$ data in the literature. Our result for $L_{10}^{\rm eff}$ agrees with \cite{DHG98,DS}, but our estimation includes a more careful assessment of the theoretical errors. The $\rm{3.2\,\sigma}$ discrepancy between the estimation of ref.~\cite{NAR01} and ours is caused by an underestimation of the systematic error associates with the duality-point approach used in that reference. In \cite{DS} also $C_{87}^{\rm eff}$ is determined with a good agreement with our result again. The extraction of $L_{10}^r(\mu)$ from $L_{10}^{\rm eff}$ has only been done previously in ref.~\cite{DHG98}, at $\cO(p^4)$.

Our determinations of $L_{10}^r(M_\rho)$ and $C_{87}^r(M_\rho)$ agree within errors with the large--$N_C$ estimates based on lowest-meson dominance \cite{KN01,CEE04,ABT00,PI02} $L_{10} \approx -3 f_\pi^2 / (8 M_V^2) \approx -5.4\cdot 10^{-3}$ and $C_{87} \approx 7 f_\pi^2 / (32 M_V^4) \approx 5.3\cdot 10^{-3}~\mathrm{GeV}^{-2}$ and with the result of ref. \cite{MP08} for $C_{87}$, based on Pad\'e Approximants. These predictions, however, are unable to fix the scale dependence which is of higher-order in $1/N_C$. More recently, the resonance chiral theory Lagrangian \cite{CEE04,EGPdR89} has been used to analyse the correlator $\Pi(s)$ at NLO order in the $1/N_C$ expansion. Matching the effective field theory description with the short-distance QCD behaviour, the two LECs are determined, keeping full control of their $\mu$ dependence. The theoretically predicted values $L_{10}^r(M_\rho) = -(4.4 \pm 0.9) \cdot 10^{-3}$ and $C_{87}^r(M_\rho)=(3.6 \pm 1.3) \cdot 10^{-3}$  GeV$^{-2}$ \cite{PRS08} are in perfect agreement with our determinations, although less precise. A recent lattice estimate \cite{LAT08} finds $L_{10}^r(M_\rho) = -(5.2 \pm 0.5) \cdot 10^{-3}$ at order $p^4$, in good agreement with our result (\ref{valL10p4}).

Using the results of ref.~\cite{GHI07}, the SU(2) $\chi PT$ LEC $\overline l_5$ can be extracted from $L_{10}^r(\mu)$. We find $\rm{\overline l_5 = 13.30 \pm 0.11}$ at $\cO(p^4)$ and $\overline l_5 = 12.24 \pm 0.21$ at $\cO(p^6)$.

Recent analyses of the decay $\pi^+ \!\to\! l^+ \nu \gamma$ at $\cO(p^6)$ have provided accurate values for the combinations $L_9\!+\!L_{10}$ \cite{UP08} and $\overline l_5\! -\! \overline l_6$ \cite{BT97}, that can be combined with our results to get $L_9^r(M_\rho)=(5.5 \pm 0.4) \cdot 10^{-3}$ and $\overline l_6 = 15.22 \pm 0.39$ to order $p^6$, that are in perfect agreement with refs.~\cite{BT02,BCT98}.

\section*{Acknowledgements}
M. G.-A. is indebted to MICINN (Spain) for a FPU Fellowship. Work partly supported by the EU network FLAVIAnet [MRTN-CT-2006-035482], by MICINN, Spain [FPA2007-60323, FPA2006-05294 and CSD2007-00042 --CPAN--] and by Junta de Andaluc\'{\i}a [Grants P05-FQM 191, P05-FQM 467 and P07-FQM 03048].

\end{document}